\documentclass[twocolumn,preprintnumbers,amsmath,amssymbm,prd]{revtex4}
\usepackage{epsfig}
\usepackage{graphicx}

\begin{document}
\title{Onset of superradiant instabilities in the hydrodynamic vortex model}
%\title{Analytic treatment of the hydrodynamic vortex: the onset of superradiant instabilities}
\author{Shahar Hod}
\affiliation{The Ruppin Academic Center, Emeq Hefer 40250, Israel}
\affiliation{ } \affiliation{The Hadassah Institute, Jerusalem
91010, Israel}
\date{\today}

\begin{abstract}
\ \ \ The hydrodynamic vortex, an effective spacetime geometry for
propagating sound waves, is studied {\it analytically}. In contrast
with the familiar Kerr black-hole spacetime, the hydrodynamic vortex
model is described by an effective acoustic geometry which has no
horizons. However, this acoustic spacetime possesses an ergoregion,
a property which it shares with the rotating Kerr spacetime. It has
recently been shown numerically that this physical system is
linearly unstable due to the superradiant scattering of sound waves
in the ergoregion of the effective spacetime. In the present study
we use analytical tools in order to explore the onset of these
superradiant instabilities which characterize the effective
spacetime geometry. In particular, we derive a simple analytical
formula which describes the physical properties of the hydrodynamic
vortex system in its critical (marginally-stable) state, the state
which marks the boundary between stable and unstable fluid
configurations. The analytically derived formula is shown to agree
with the recently published numerical data for the hydrodynamic
vortex system.
\end{abstract}
\bigskip
\maketitle

%]

\section{Introduction}

One of the most remarkable characteristics of the rotating Kerr
black-hole spacetime \cite{Kerr} is the existence of an {\it
ergoregion} \cite{Chan}: a region in which all physical objects must
co-rotate with the spinning black hole. In particular, no physical
observer inside the ergoregion can remain at rest with respect to
inertial asymptotic observes.

The presence of the ergoregion in the Kerr black-hole spacetime is
responsible for the intriguing phenomenon of {\it superradiant} wave
scattering: it was first realized by Zel'dovich \cite{Zel} (see also
\cite{PressTeu1,PressTeu2}) that a co-rotating bosonic field of the
form $e^{im\phi}e^{-i\omega t}$ interacting with a spinning Kerr
black hole of angular velocity $\Omega$ can be amplified (gain
energy) if the incident wave field satisfies the superradiant
condition
\begin{equation}\label{Eq1}
\omega<m\Omega\  .
\end{equation}

The superradiant scattering of bosonic fields in the Kerr black-hole
ergoregion (the extraction of rotational energy from the black hole)
has the potential to destabilize the spacetime geometry \cite{Cars}.
However, the Kerr black hole is known to be stable against massless
perturbation fields \cite{PressTeu2,Whit}. The stability of the Kerr
spacetime against the superradiant scattering of massless bosonic
fields in its ergoregion may be attributed to the absorption
properties of the black-hole horizon \cite{Cars}. In particular, the
black-hole horizon acts as a one-way membrane which absorbs the
(potentially dangerous) perturbation fields before any instability
has the chance to develop in the ergoregion
\cite{Notebom,Ins1,Ins2,Ins3}.

The reasoning presented above \cite{Cars} suggests that {\it
horizonless} spacetimes which possess ergoregions may generally be
unstable to superradiant scattering of bosonic fields in their
ergoregions. This suggestive argument was raised long ago by
Friedman \cite{Fri}. Most recently, Oliveira et. al. \cite{Cars}
have explored analogous ergoregion instabilities which may develop
in fluid flow dynamics. It was first shown by Unruh \cite{Unr} that
the characteristic wave equation for sound waves propagating inside
fluids is analogous to the Klein-Gordon wave equation for massless
scalar fields propagating in curved spacetimes [see Eq. (\ref{Eq7})
below].

Oliveira et. al. \cite{Cars} have studied numerically the {\it
hydrodynamic vortex} model, a two-dimensional purely circulating
flow of a vorticity free ideal fluid. This acoustic system is
analogous to an effective horizonless spacetime which nevertheless
possesses an ergoregion \cite{Sla}. In accord with the arguments
presented in \cite{Fri}, it has been established in \cite{Cars} that
the effective spacetime geometry which corresponds to the
hydrodynamic vortex system is characterized by unstable acoustic
perturbation modes. This is an ergoregion instability \cite{Cars}
which is related to the absence of an event horizon in the effective
rotating spacetime.

A remarkable feature of the hydrodynamic vortex system is the
existence, for each given value of the sound mode harmonic index $m$
[see Eq. (\ref{Eq8}) below], of {\it marginally-stable} fluid
configurations. These stationary configurations mark the boundary
%(for each given value of the sound mode index $m$)
between stable and unstable fluid flows. The main goal of the
present study is to obtain an {\it analytical} formula which
describes the physical properties of the hydrodynamic vortex system
in its critical (marginally-stable) state.

\section{Description of the system}

We study the dynamics of a vorticity free barotropic ideal fluid.
Assuming a two-dimensional purely circulating flow in the $xy$
plane, the background (unperturbed) fluid velocity is characterized
by \cite{Cars}
\begin{equation}\label{Eq2}
v_r=v_z=0\ \ \ ; \ \ \ v_{\phi}=v_{\phi}(r)\  .
\end{equation}
Here $r$ and $\phi$ are respectively the radial and azimuthal
coordinates in the $xy$ plane, and $z$ is the coordinate
perpendicular to the plane of flow.

Irrotationality of the fluid flow (vorticity free flow) implies that
the tangential component of the velocity field, $v_{\phi}$, is given
by \cite{Cars}
\begin{equation}\label{Eq3}
v_{\phi}={C\over r}\\ ,
\end{equation}
where the constant $C$ characterizes the circulation strength of the
fluid. Conservation of angular momentum yields
\begin{equation}\label{Eq4}
\rho v_{\phi}r=\text{const.}
\end{equation}
which, together with Eq. (\ref{Eq3}), implies that the fluid
background density $\rho$ is constant. The assumption of a
barotropic fluid then implies that the background pressure $P$ and
the speed of sound $c$ are also constants.

The two-dimensional circulating fluid flow produces an effective
acoustic spacetime, known as the hydrodynamic vortex
\cite{Cars,Sla,Fed,Dol}, which is characterized by the non-trivial
\cite{Notetri} line element
%\cite{Cars,Sla,Fed,Dol}
\begin{equation}\label{Eq5}
ds^2=-c^2\Big(1-{{C^2}\over{c^2r^2}}\Big)dt^2+dr^2-2Cdtd\phi+r^2d\phi^2+dz^2\
.
\end{equation}

The rotating acoustic spacetime geometry (\ref{Eq5}) possesses an
ergoregion whose outer boundary is determined by the circle at which
the fluid flow velocity, $|C|/r$, equals the speed of sound $c$
\cite{Cars,Sla,Fed,Dol}:
\begin{equation}\label{Eq6}
r_{\text{ergo}}={{|C|}\over{c}}\  .
\end{equation}
(We shall henceforth use units in which $c=1$. Note that in these
units $C$ has the dimensions of length).

We shall now consider small perturbations to the background fluid
flow. These perturbations (sound waves) propagate in the acoustic
spacetime and their linearized Navier-Stokes dynamics is governed by
the Klein-Gordon wave equation \cite{Unr,Cars,Notefl}
\begin{equation}\label{Eq7}
\nabla^\nu\nabla_{\nu}\Psi={{1}\over{\sqrt{|g|}}}\partial_{\mu}\Big(\sqrt{|g|}
g^{\mu\nu}\partial_{\nu}\Psi\Big)=0\  .
\end{equation}

It proves useful to decompose the perturbation field $\Psi$ in the
form \cite{Notecy}
\begin{equation}\label{Eq8}
\Psi(t,r,\phi,z)={{1}\over{\sqrt{r}}}\sum_{m=-\infty}^{\infty}\psi_m(r;\omega)e^{im\phi}e^{-i\omega
t}\ ,
\end{equation}
The $\phi$-periodicity of the angular function $e^{im\phi}$ enforces
the azimuthal harmonic index $|m|$ to be an integer. Substituting
the decomposition (\ref{Eq8}) into the Klein-Gordon wave equation
(\ref{Eq7}), one obtains the characteristic radial equation
\begin{equation}\label{Eq9}
\Big[{{d^2}\over{dr^2}}+\Big(\omega-{{Cm}\over{r^2}}\Big)^2-{{m^2-{1\over
4}}\over{r^2}}\Big]\psi_m(r;\omega)=0\
\end{equation}
for each field mode \cite{Notemm}. (We shall henceforth omit the
harmonic index $m$ for brevity).

\section{Boundary conditions}

The background velocity field (\ref{Eq3}) is singular at the origin,
signaling a breakdown of the physical description. In order to mimic
a possible experimental scenario in the laboratory, Oliveira et. al.
\cite{Cars} have suggested to impose physically acceptable boundary
conditions at a {\it finite} radial location, $r=r_{\text{min}}$. In
particular, it was assumed in \cite{Cars} that an infinitely long
cylinder of radius $r_{\text{min}}$ made of a certain material with
acoustic impedance $Z$ \cite{Lax} is placed at the center of the
fluid system.

Oliveira et. al. \cite{Cars} considered two types of boundary
conditions (BCs) at the surface $r=r_{\text{min}}$ of the central
cylinder, characterizing  two limiting values of the cylinder
acoustic impedance: Extremely low-Z materials \cite{Lax} are
characterized by the Dirichlet-type boundary condition \cite{Cars}:
\begin{equation}\label{Eq10}
\psi(r=r_{\text{min}})=0\ \ \ , \ \ \ \text{BCI}\  ,
\end{equation}
whereas extremely high-Z materials \cite{Lax} (that is, a very rigid
boundary cylinder \cite{Cars}) are characterized by the Neumann-type
boundary condition \cite{Cars}:
\begin{equation}\label{Eq11}
{{d\Psi}\over{dr}}(r=r_{\text{min}})={{d(\psi/\sqrt{r})}\over{dr}}(r=r_{\text{min}})=0\
\ \ , \ \ \ \text{BCII}\  .
\end{equation}
Following \cite{Cars}, we shall consider purely outgoing waves at
large distances from the cylinder:
\begin{equation}\label{Eq12}
\psi(r\to\infty)\sim e^{i\omega r}\  .
\end{equation}

\section{The ergoregion instability of the hydrodynamic vortex}

As emphasized above, the instability of the hydrodynamic vortex
system studied in \cite{Cars} is closely related to the phenomenon
of superradiant scattering \cite{Sla} of sound waves in the
ergoregion of the effective spacetime geometry (\ref{Eq5}). Thus,
the simple inequality \cite{Cars}
\begin{equation}\label{Eq13}
r_{\text{min}}<C\
\end{equation}
acts as a necessary requirement for the triggering of the ergoregion
instability. It simply states that the ergoregion [whose outer
boundary is given by $r=r_{\text{ergo}}=C$, see Eq. (\ref{Eq6})]
must be part of the physical system.

It should be emphasized, however, that not every hydrodynamic vortex
system with $r_{\text{min}}<r_{\text{ergo}}=C$ is unstable under
perturbations of the $m$th sound mode \cite{Cars}. In particular, a
remarkable feature of the hydrodynamic vortex system is the
existence, for each given set $(C,m)$ of the fluid and field
parameters, of a critical cylinder radius,
\begin{equation}\label{Eq14}
r_{\text{min}}=r^{*}_{\text{min}}(C,m)\  ,
\end{equation}
which supports stationary ({\it marginally-stable}) fluid
configurations.

The critical (maximal) cylinder radius (\ref{Eq14}) marks the
boundary between stable and unstable composed fluid-cylinder
configurations: composed systems whose cylinder radius lies in the
regime $r_{\text{min}}>r^{*}_{\text{min}}(C,m)$ are stable under
perturbations of the $m$th sound mode (that is, the sound mode
decays in time), whereas composed systems whose cylinder radius lies
in the regime $r_{\text{min}}<r^{*}_{\text{min}}(C,m)$ are unstable
under perturbations of the $m$th sound mode (that is, the sound mode
grows exponentially over time).

\section{Onset of the ergoregion instability in the hydrodynamic vortex}

The boundary conditions (\ref{Eq10})-(\ref{Eq12}) single out a
discrete set of complex resonances
$\{\omega_n(C,m,r_{\text{min}})\}$ \cite{Notenn}. These quasinormal
resonances characterize the temporal response of the hydrodynamic
system to external (sound mode) perturbations. Note, in particular,
that stable (exponentially suppressed) sound modes are characterized
by $\Im\omega<0$, whereas unstable (growing in time) sound modes are
characterized by $\Im\omega>0$. The stationary (marginally-stable)
resonances, which are the solutions we shall be interested in in
this study, are characterized by $\Im\omega=0$.

In this section we shall analyze the marginally-stable
($\Im\omega=0$) resonances of the hydrodynamic vortex system. As we
shall show below, this hydrodynamic fluid system is actually
characterized by genuine {\it static} resonances with
\begin{equation}\label{Eq15}
\Re\omega=\Im\omega=0\  .
\end{equation}
In particular, we shall now find the {\it discrete} set of critical
cylinder radii, $\{r^{*}_{\text{min}}(C,m;n)\}$, which support these
marginally-stable fluid configurations.

Remarkably, the radial equation (\ref{Eq9}) can be solved {\it
analytically} for the marginally-stable modes (\ref{Eq15}). The
general solution of Eq. (\ref{Eq9}) with $\omega=0$ can be expressed
in terms of the Bessel functions of the first and second kinds (see
Eq. 9.1.53 of \cite{Abram}):
\begin{equation}\label{Eq16}
\psi(r;\omega=0)=Ar^{1\over 2}J_m\Big({{Cm}\over{r}}\Big)+Br^{1\over
2}Y_{m}\Big({{Cm}\over{r}}\Big)\ ,
\end{equation}
where $A$ and $B$ are normalization constants. The asymptotic
large-$r$ limit ($Cm/r\to 0$) of Eq. (\ref{Eq16}) is given by (see
Eqs. 9.1.7 and 9.1.9 of \cite{Abram})
\begin{eqnarray}\label{Eq17}
\psi(r\to\infty;\omega=0)&=&{{A}\over{m!}}\Big({{Cm}\over{2}}\Big)^mr^{-m+{1\over2}}
\nonumber \\ &&
-{{B(m-1)!}\over{\pi}}\Big({{Cm}\over{2}}\Big)^{-m}r^{m+{1\over2}}\
. \nonumber \\ &&
\end{eqnarray}
A physically acceptable (finite) solution at infinity requires
$B=0$, which implies
\begin{equation}\label{Eq18}
\psi(r;\omega=0)=Ar^{1\over 2}J_m\Big({{Cm}\over{r}}\Big)\ .
\end{equation}

Taking cognizance of the boundary conditions
(\ref{Eq10})-(\ref{Eq11}) which characterize the acoustic properties
of the central cylinder, one finds that the marginally-stable
resonances (\ref{Eq15}) of the hydrodynamic vortex system correspond
to the following {\it discrete} radii of the cylinder:
\begin{equation}\label{Eq19}
r^{*}_{\text{min}}(C,m;n)={{Cm}\over{j_{m,n}}} \ \ \ , \ \ \
\text{BCI}\  ,
\end{equation}
and
\begin{equation}\label{Eq20}
r^{*}_{\text{min}}(C,m;n)={{Cm}\over{j^{'}_{m,n}}} \ \ \ , \ \ \
\text{BCII}\  ,
\end{equation}
where $n=1,2,3,...\ $. Here $j_{m,n}$ is the $n$th positive zero of
the Bessel function $J_m(x)$ and $j^{'}_{m,n}$ is the $n$th positive
zero of its first spatial derivative $J^{'}_m(x)$. The real zeros of
these functions were studied by many authors, see e.g.
\cite{Abram,Bes}.

For large values of the acoustic harmonic index, $m\gg n$, one may
use the asymptotic relations (see Eqs. 9.5.14 and 9.5.16 of
\cite{Abram}) $j_{m,n}=m[1+b_n m^{-2/3}+O(m^{-4/3})]$ and
$j^{'}_{m,n}=m[1+b^{'}_n m^{-2/3}+O(m^{-4/3})]$ \cite{Notebb}.
Substituting these relations into Eqs. (\ref{Eq19}) and
(\ref{Eq20}), one finds
\begin{equation}\label{Eq21}
r^{*}_{\text{min}}(C,m\gg n;n)=C[1-b_n m^{-2/3}+O(m^{-4/3})] \ \ \ ,
\ \ \ \text{BCI}\ ,
\end{equation}
and
\begin{equation}\label{Eq22}
r^{*}_{\text{min}}(C,m\gg n;n)=C[1-b^{'}_n m^{-2/3}+O(m^{-4/3})] \ \
\ , \ \ \ \text{BCII}\ .
\end{equation}
It is worth emphasizing that Eqs. (\ref{Eq21})-(\ref{Eq22}) provides
an analytical quantitative explanation for the asymptotic large-$m$
behavior of the hydrodynamic vortex system as numerically presented
in Fig. 5 of \cite{Cars}.

For large overtone numbers, $n\gg m$, one may use the asymptotic
relations (see Eqs. 9.5.12 and 9.5.13 of \cite{Abram})
$j_{m,n}=(n+m/2-1/4)\pi+O(m^2/n)$ and
$j^{'}_{m,n}=(n+m/2-3/4)\pi+O(m^2/n)$. Substituting these relations
into Eqs. (\ref{Eq19}) and (\ref{Eq20}), one finds
\begin{equation}\label{Eq23}
r^{*}_{\text{min}}(C,m;n\gg m)={{Cm}\over{\pi n}}[1+O(m/n)]\
\end{equation}
for both types of boundary conditions. The relation (\ref{Eq23})
implies that, large overtone modes must be supported by small radii
cylinders in order to be able to trigger superradiant instabilities
in the hydrodynamic vortex system.

\section{Analytical vs. numerical results}

We shall now compare the predictions of the analytically derived
formulas (\ref{Eq19})-(\ref{Eq20}) for the critical cylinder radii,
$r^{*}_{\text{min}}(C,m;n)$, with the corresponding numerical data
recently published by Oliveira et. al. \cite{Cars}. In Table
\ref{Table1} we present such a comparison, from which one finds a
remarkably excellent agreement between the analytical formulas
(\ref{Eq19})-(\ref{Eq20}) and the numerical results of \cite{Cars}.

\begin{table}[htbp]
\centering
\begin{tabular}{|c|c|c|c|}
\hline \ \ BC \ \ & \ $m$\ \ & \ \
$r^{*}_{\text{min}}(\text{Numerical})$\ \ \ & \ \
$r^{*}_{\text{min}}(\text{Analytical})$\ \ \ \\
\hline
\ \ I \ \ & \ \ 1 \ \ & \ \ 0.13\ \ \ & \ \ 0.1305\ \ \ \\
\ \ I \ \ & \ \ 2 \ \ & \ \ 0.19\ \ \ & \ \ 0.1947\ \ \ \\
\ \ II \ \ & \ \ 1 \ \ & \ \ 0.27\ \ \ & \ \ 0.2716\ \ \ \\
\ \ II \ \ & \ \ 2 \ \ & \ \ 0.33\ \ \ & \ \ 0.3274\ \ \ \\
\hline
\end{tabular}
\caption{Marginally stable resonances of the hydrodynamic vortex
system. We display the critical cylinder radii,
$r^{*}_{\text{min}}$, for the fundamental $n=1$ sound mode with
fluid circulation $C=0.5$ as obtained from the analytically derived
formulas (\ref{Eq19}) and (\ref{Eq20}). We also display the
corresponding critical radii as extracted from the numerical data
presented in \cite{Cars}. One finds a remarkably good agreement
between the analytical formulas (\ref{Eq19})-(\ref{Eq20}) and the
numerically computed \cite{Cars} critical cylinder radii.}
\label{Table1}
\end{table}

\section{Summary and discussion}

In this paper, we have used analytical tools in order to analyze the
{\it marginally-stable} resonances of the hydrodynamic vortex
system, an effective spacetime geometry for sound waves. These
resonances are fundamental to the physics of sound waves in the
hydrodynamic acoustic spacetime: in particular, they mark the onset
of the superradiant instability in the hydrodynamic vortex system.

In order to mimic a possible experimental scenario in the
laboratory, Oliveira et. al. \cite{Cars} have recently suggested to
place a long cylinder of radius $r_{\text{min}}$ at the center of
the fluid system, on which physically acceptable boundary conditions
[see Eq. (\ref{Eq10}) and (\ref{Eq11})] are imposed. A remarkable
feature of the hydrodynamic vortex system is the existence, for each
given set $(C,m)$ of the fluid and sound mode parameters, of a
critical cylinder radius, $r_{\text{min}}=r^{*}_{\text{min}}(C,m)$,
which supports marginally-stable fluid configurations.

In the present study we have derived the characteristic resonance
conditions [see Eq. (\ref{Eq19}) and (\ref{Eq20})] for these
marginally-stable composed fluid-cylinder configurations. In
particular, it was shown that the critical cylinder radius
$r^{*}_{\text{min}}$ (the cylinder radius which, for given
parameters $C$ and $m$ of the system, marks the boundary between
stable and unstable fluid-cylinder configurations) can be expressed
in terms of the simple zeros of the Bessel function and its first
derivative.

It was shown that the {\it analytically} derived formulas for the
critical cylinder radius [see Eq. (\ref{Eq19}) and (\ref{Eq20})]
agree with the recently published {\it numerical} data for the
hydrodynamic vortex system.

Finally, it is worth emphasizing that the physical significance of
the critical cylinder radius, $r^{*}_{\text{min}}$, lies in the fact
that it is the {\it outermost} location of the cylinder which allows
the extraction of rotational energy from the circulating fluid (from
the ergoregion of the effective spacetime geometry).

\bigskip
\noindent
{\bf ACKNOWLEDGMENTS}
\bigskip

This research is supported by the Carmel Science Foundation. I thank
Yael Oren, Arbel M. Ongo and Ayelet B. Lata for stimulating
discussions.

%\newpage


\begin{thebibliography}{99}
\bibitem{Notebom} It is well known that,
in order to trigger exponentially growing superradiant instabilities
in the rotating Kerr black-hole spacetime, some additional mechanism
is required in order to confine the extracted energy in the vicinity
of the black-hole horizon. Several such confinement mechanisms are
known: (1) An artificial reflecting mirror which is placed around
the black hole \cite{PressTeu1,Ins1}, (2) A massive bosonic field
whose mass term effectively acts as a mirror \cite{Ins2}, and (3)
Non asymptotically flat (e.g., an Anti-de Sitter) boundary to the
spacetime \cite{Ins3}.
\bibitem{Notetri} Note that a background fluid with zero circulation
($C=0$) is characterized by a trivial flat space line element.
\bibitem{Notecy} That is, we consider cylindrical wave modes.
\bibitem{Notemm} Note that the radial wave equation
(\ref{Eq9}) is invariant under the symmetry transformation $m\to -m$
with $C\to -C$. Thus, one can assume without loss of generality that
$m>0$ and $C>0$.
\bibitem{Notebb} The coefficients $b_n$ and $b^{'}_n$ are determined by the
negative zeros of the Airy functions, see \cite{Abram} for details.
The values of these constants are given in \cite{Abram}.
\end{thebibliography}
\end{document}